# Pose correction scheme for camera-scanning Fourier ptychography based on camera calibration and homography transform

**Baiqi Cui**
School of Optics and Photonics,
Beijing Institute of Technology
Beijing 100081, China
cuibaiqi@126.com

**Shaohui Zhang**
School of Optics and Photonics,
Beijing Institute of Technology
Beijing 100081, China
Yangtze Delta Region Academy of Beijing Institute of Technology, China
zhangshaohui@bit.edu.cn

**Yechao Wang**
Beijing Institute of Space Mechanics & Electricity,
Beijing 100094, China
cast_wangyc_508@163.com

**Yao Hu**
School of Optics and Photonics,
Beijing Institute of Technology
Beijing 100081, China
huy08@bit.edu.cn

**Qun Hao**
School of Optics and Photonics,
Beijing Institute of Technology
Beijing 100081, China
Yangtze Delta Region Academy of Beijing Institute of Technology, China
qhao@bit.edu.cn

## Abstract

Fourier ptychography (FP), as a computational imaging method, is a powerful tool to improve imaging resolution. Camera-scanning Fourier ptychography extends the application of FP from micro to macro creatively. Due to the non-ideal scanning of the camera driven by the mechanical translation stage, the pose error of the camera occurs, greatly degrading the reconstruction quality, while a precise translation stage is expensive and not suitable for wide-range imaging. Here, to improve the imaging performance of camera-scanning Fourier ptychography, we propose a pose correction scheme based on camera calibration and homography transform approaches. The scheme realizes the accurate alignment of data set and location error correction in the frequency domain. Simulation and experimental results demonstrate this method can optimize the reconstruction results and realize high-quality imaging effectively. Combined with the feature recognition algorithm, the scheme provides the possibility for applying FP in remote sensing imaging and space imaging.

## 1 Introduction

Improving the resolution of optical imaging systems and recovering the details of objects have always been the research hotspot in the field of advanced optical remote sensing and high-performance imaging. The diffraction limit of traditional optical systems is a core factor restricting the resolution improvement of long-distance imaging systems, the image with diffraction blur provides limited information of the object. In general, increasing the aperture size of the pupil is the most direct way to improve the resolution of the optical system. Nevertheless, as the diameter of the lens aperture increases, the weight, volume, and cost of the optical system increase dramatically [1,2]. Therefore, the aperture size cannot be infinitely increased in the theory of traditional incoherent imaging. At the same time, affected by the non-ideal lens manufacturing process and processing environment, there are inevitable manufacturing errors and profile deformations in actual optical systems. These problems will be reflected as the wavefront aberration at the pupil of the entire optical imaging system, making it difficult for the image quality to reach the diffraction limit.

As a concise and efficient super-resolution imaging scheme, FP has attracted much attention since it was proposed by Zheng *et al.* [3] in 2013. Combining thoughts of synthetic aperture [4–13] and phase retrieval [14–20], FP realizes wide-field and high-resolution imaging simultaneously by using computational methods and different implementations [21]. FP was only used in microscopic imaging when it was first proposed, called Fourier ptychography microscopy (FPM), which replaced the light source of a standard microscope with an LED to provide varying-angle illumination. Due to the equivalence relationship of varying-angle illumination in space and translation in the frequency domain, different information of sample in the frequency domain can be recorded by lighting different LED sequentially. Then the alternating projection algorithm [14,22] will be applied, which iteratively imposes constraints in the spatial domain and frequency domain to recover the complex amplitude of the sample.

In recent years, FP has been extended from the initial microscopic imaging method into a more general technique for macroscopic imaging [1,23-26], aberration correction [27], and 3D imaging [28], *etc.* For example, Dong *et al.* [23] proposed a scheme called camera-scanning Fourier ptychography. This scheme extends the application of FP from micro to macro. Since the angle of the illumination beam is difficult to modulate in a wide range in macroscopic imaging, shifting each sub-aperture in the frequency domain cannot be realized by changing the light source location [29]. In this scheme, the camera aperture is placed on the Fourier plane (far-field Fraunhofer diffraction plane) of the sample and scanning laterally in this plane, and a series of low resolution (LR) images similar to those in FPM that correspond to different sub-spectrum regions can be obtained. Similar to FPM, by applying the phase retrieval scheme to the data set of these LR images, far-field super-resolution imaging can be achieved finally. The reconstruction result of this scheme is determined by the way the complex wavefront leaves the sample rather than the way it enters the sample. Therefore, the thickness of the sample does not affect the reconstruction result.

Camera-scanning Fourier ptychography keeps the advantages of FP, which are simplicity, cheapness, and efficiency. It has a significant effect on resolution improvement, showing strong competitiveness in the field of far-field imaging. Nevertheless, because of the non-ideal mechanical movement of the camera, the difference between the actual system parameters and the ideal system parameters is more significant than that in FPM systems [30,31]. Two of the most important factors degrading the reconstruction result are the difference in the LR data set caused by the change of camera pose and the non-corresponding of the two constraints in the alternating projection algorithm. Nevertheless, these inevitable scanning errors, including non-ideal translation and twist of the camera, are random and appear during every scanning process. Using a concise and effective method to extract and eliminate these errors is significant for using FP in far-field macro imaging.

The scanning of the camera is a process of pose transformation. At the same time, the sample is fixed, and the relative pose transformation of the two is closely related to the camera calibration process. Camera calibration is widely used in the field of machine vision with the purpose to establish a transform function model from a 3D space to a 2D image [32-38]. As intermediate products of camera calibration, the rotation and translation matrices between the camera and the sample at different scanning locations can be obtained simultaneously. Therefore, by adding feature points artificially and introducing camera calibration into the camera-scanning Fourier ptychography process, we can obtain different scanning location information of different scanning processes. Besides, the twist of the camera can be corrected by recognizing the feature points to implement the homography transform, finally, the camera pose can be corrected.

In this paper, a camera pose correction scheme based on camera calibration and homography transform is proposed. The imaging model of FP, image alignment principle, and location error correction method are described. The simulation results prove the feasibility of this scheme. The experimental results effectively demonstrate the proposed scheme can improve the precision of camera-scanning far-field Fourier ptychography and achieve high-quality reconstruction.

## 2  Principle and method

### 2.1 The physical model of camera-scanning Fourier ptychography

The smaller object has more high-frequency information. If one wants to increase the resolution, recording more frequency information is necessary. The camera aperture in camera-scanning Fourier ptychography system is placed on the Fourier plane of the sample to realize the expansion of the aperture and the recording of high-frequency information. Combined with the alternating projection algorithm, the scheme can realize super-resolution imaging.

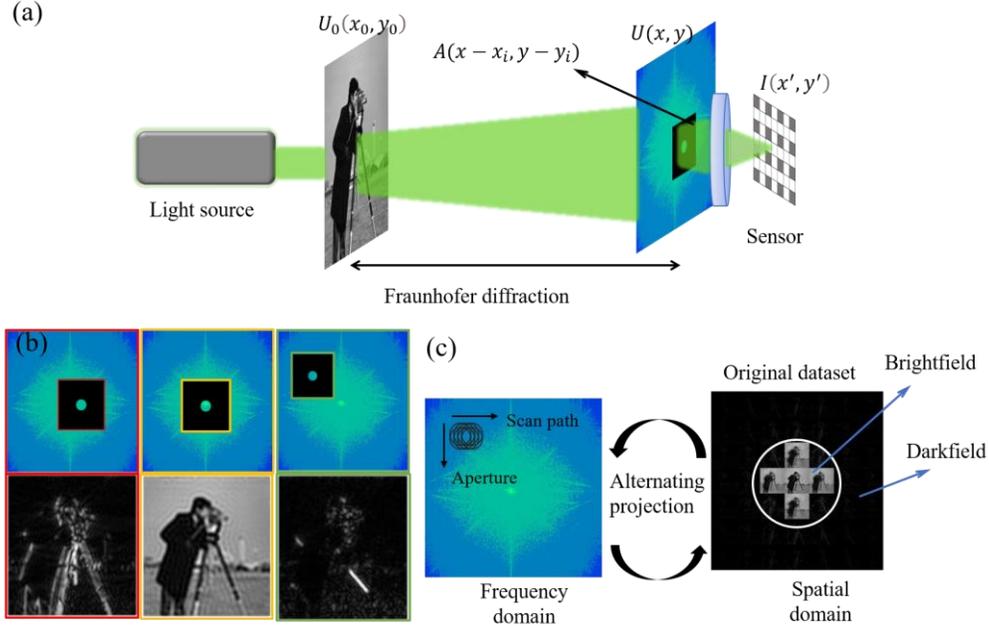

Figure 1: The imaging principle of camera-scanning Fourier ptychography system. (a) The schematic of the setup. (b) Specific sub-regions of the spectrum and their corresponding intensity image. (c) shows the alternating projection with constraints in the frequency domain and spatial domain. Besides, the data set and the division of brightfield and darkfield are displayed.

The imaging principle of the camera-scanning Fourier ptychography system is shown in Figure 1(a), which uses a light source with enough coherence [23]. Figure 1(b) shows the intensity images corresponding to different scanning regions of the Fourier plane, and obviously, a specific sub-region of the spectrum corresponds to a specific intensity image. The complex amplitude distribution obtained after far-field Fraunhofer diffraction can be expressed as

$$U(x,y) = \frac{e^{j\lambda z} e^{j\frac{k}{2z}(x^2+y^2)}}{j\lambda z} \iint_{\infty} U_0(x_0,y_0) \exp[-j\frac{2\pi}{\lambda z}(xx_0+yy_0)]dx_0 dy_0 = \frac{e^{j\lambda z} e^{j\frac{k}{2z}(x^2+y^2)}}{j\lambda z} \mathcal{F}_{1/\lambda z}\{U_0(x_0,y_0)\} \quad (1)$$

where $U_0(x_0,y_0)$ is the object light wave function on the back surface of the illuminated sample, $(x_0,y_0)$ is the coordinate of object plane, $\lambda$ is the illumination wavelength, $k=2\pi/\lambda$ is the wave number, $\mathcal{F}$ represents the two-dimension Fourier transform, $1/\lambda z$ is a scale factor, and $(x,y)$ is the coordinate of the Fourier plane.

Here, the intensity of the sensor plane is deduced by taking a single scanning as an example. After the low-pass filtering of the aperture, the light wave field distribution on the back surface of the aperture is

$$U'(x,y) = U(x,y)A(x-x_i, y-y_i) \quad (2)$$

where $A(x-x_i, y-y_i)$ is the aperture function of the $i^{th}$ shot with a center $(x_i, y_i)$. Ignoring the constant phase term and the scale factor, the intensity distribution of the sensor plane can be written as the Fourier transform form [1]

$$I_i(x', y') = |\mathcal{F}^-\{U(x,y)A(x-x_i, y-y_i)\}|^2 \qquad (3)$$

where $\mathcal{F}^-$ denotes the inverse Fourier transform, $(x', y')$ is the coordinate in the sensor plane.

The original data set obtained is shown in Figure 1(c), which can be written as

$$I = \sum_{(x_i, y_i)} |\mathcal{F}^-\{U(x,y)A(x-x_i, y-y_i)\}|^2 \qquad (4)$$

Then, as shown in Figure 1(c), the constraints in the frequency domain and spatial domain are imposed iteratively in the alternating projection algorithm loop. In general, take a single iteration as an example, the constraint in the spatial domain is the intensity of an image in the data set, and the corresponding constrain in the frequency domain is determined by the pupil size and its location. Since the camera aperture is placed on the Fourier plane, the constrain in the frequency domain is determined by the aperture size and camera location in space. Ensuring that the two constraints are matched well when reconstruction, after several loops, the resolution of the reconstruction result will be obviously improved compared to a single scanning.

Due to the introduction of mechanical parts in the scanning process, the difference between the actual system parameters and the ideal system parameters is also introduced. The two main factors are 1) the difference between images in the dataset caused by the change of the camera pose which makes that a point in space corresponds to a different pixel in different images in the data set, and 2) the difference between actual scanning step and default step, which makes the spatial constraints do not correspond to the frequency domain constraints in the reconstruction process. To realize high-quality FP imaging, it is necessary to implement the extraction and correction of camera location error and accurate alignment of original data set to correct camera pose error eventually.

## 2.2 Extraction of camera location information

The imaging process converts the 3D world into a 2D image. Let the 3D scene be the input and the 2D image be the output, the imaging process can be considered as a transform function, and the calibration process of a camera is to establish the transform function model. Using the results of camera calibration, which contain the transform matrices between the world coordinate system and pixel coordinate system, the rotation and translation of every scanning location can be solved. This does not provide the absolute pose information of the camera, but the relative pose relationship of each single scanning can be solved. To find the correct locations of the constraints in the frequency domain, extracting the location information of the camera from the known camera calibration results is necessary. The locations of the pixel coordinate system, image coordinate system, camera coordinate system, and the world coordinates system are shown as Figure 2(a), and the conversion relationship of any point in the space among systems mentioned above is shown as Figure 2(b).

The known conversion relationship between the world coordinates and pixel coordinates is [32]

$$Z_c \begin{bmatrix} u \\ v \\ 1 \end{bmatrix} = \begin{bmatrix} \frac{1}{dx} & 0 & u_0 \\ 0 & \frac{1}{dy} & v_0 \\ 0 & 0 & 1 \end{bmatrix} \begin{bmatrix} f & 0 & 0 & 0 \\ 0 & f & 0 & 0 \\ 0 & 0 & 1 & 0 \end{bmatrix} \begin{bmatrix} R & T \\ 0 & 1 \end{bmatrix} \begin{bmatrix} X_w \\ Y_w \\ Z_w \\ 1 \end{bmatrix} = MM' \begin{bmatrix} X_w \\ Y_w \\ Z_w \\ 1 \end{bmatrix} \quad (5)$$

where $M$ is the intrinsic matrix, which is an inherent property related to camera parameters, and $M'$ is the extrinsic matrix, which includes the rotation and translation transform. Rotation matrix $R$ and the translation vector $T$ represent the transformation from the world coordinate system to the camera coordinate system. The subscript $c$ represents the camera coordinate, subscript $w$ represents the world coordinate system coordinate, and $(u, v)$ represents the pixel coordinate.

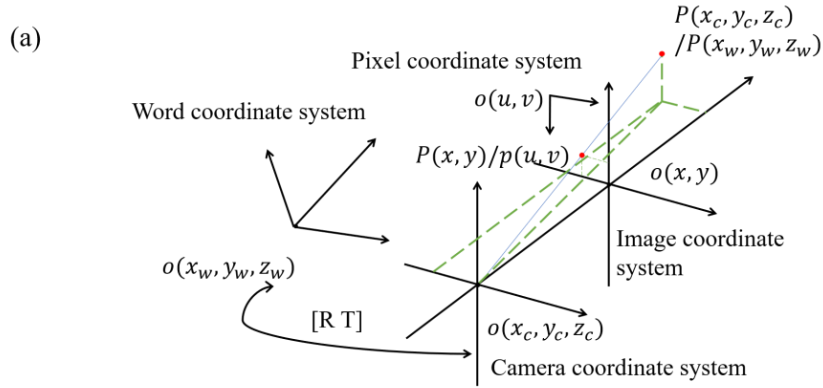

(a)

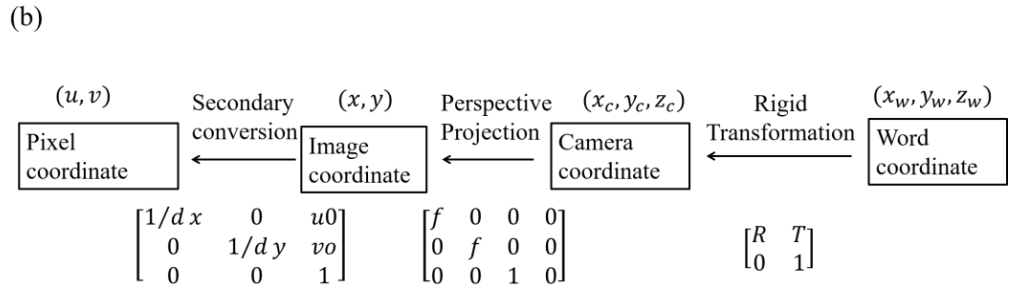

(b)

Figure 2: (a) The locations of pixel coordinate system, image coordinate system, camera coordinate system, and world coordinates system. (b) The conversion relationships of the mentioned coordinate systems.

Notice that the translation vector $T$ represents the coordinate of the origin of the word coordinate system in the camera coordinate system. When camera scanning, the scene is fixed, the translation vector provides the relative relationships of each scanning location. By calculating the pixel coordinates of the origin of the world coordinates system of each scanning location, we can obtain the pixel offset of each scanning location. Let the rotation matrix and the translation vector of the $i^{th}$ scanning be $R_i$ and $T_i$ respectively, substituting the origin $(0,0,0)$ of the world coordinate system, we obtain

$$Z_{ci}\begin{bmatrix}u_i\\v_i\\1\end{bmatrix}=M\begin{bmatrix}R_i & T_i\\0 & 1\end{bmatrix}\begin{bmatrix}0\\0\\0\\1\end{bmatrix}. \qquad(6)$$

Take the central image of brightfield as the center of the reconstruction, the location offset $(\Delta u_i, \Delta v_i)$ of each image relative to the central image can be extracted as

$$\begin{bmatrix}\Delta u_i\\\Delta v_i\end{bmatrix}=\begin{bmatrix}u_i\\v_i\end{bmatrix}-\begin{bmatrix}u_{center}\\v_{center}\end{bmatrix} \qquad(7)$$

Where $(\Delta u_i, \Delta v_i)$ is the pixel coordinate of the center image. Generally, $(\Delta u_i, \Delta v_i)$ can be used to align the original dataset when the camera does not twist, which is the rotation around the three axes of the camera coordinate system, but the twist also causes the pixel drift. If the location correction is performed individually, pixels can only be roughly aligned, which will also cause the deterioration of the reconstructed image eventually.

## 2.3 Image alignment

Shown as Eqn. 5,

$$Z_c\begin{bmatrix}u\\v\\1\end{bmatrix}=MM'\begin{bmatrix}X_w\\Y_w\\Z_w\\1\end{bmatrix}. \qquad(8)$$

Define $H = MM'$ where $H$ is the homography matrix used to describe the correspondence of points between two planes in space [32]. The homography transform is a 2D projection transform that maps points in one plane to another, which meets the requirements of camera-scanning Fourier ptychography to align the images in the original data set point-to-point. The mapping of the points between the plane of 3D space and the coordinate plane is also a homography transform [31]. Take the homography transform between two images in the original data set as an example, let $H_1$ be the homography matrix from a specific plane $\Pi_1$ in the space to the corresponding pixel plane $P_1$, $H_2$ be the homography matrix from another plane $\Pi_2$ in the space to the pixel plane $P_2$, $H_3$ be the homography matrix from $P_2$ to $P_1$, $H_4$ be the homography matrix from plane $\Pi_2$ to $\Pi_1$.

Figure 3 shows the corresponding relationships among $P_2$, $P_1$, $\Pi_1$ and $\Pi_2$. Regardless of the constant factor, the homography matrix is unique, the relations of these planes form a closed loop by these homography matrixes. Therefore, besides the direct transformation, there is another path from $P_2$ to $P_1$ using the homography transform, which can be written as

$$P_2 \to P_1 = P_2 \to \Pi_2 \to \Pi_1 \to P_1. \qquad (9)$$

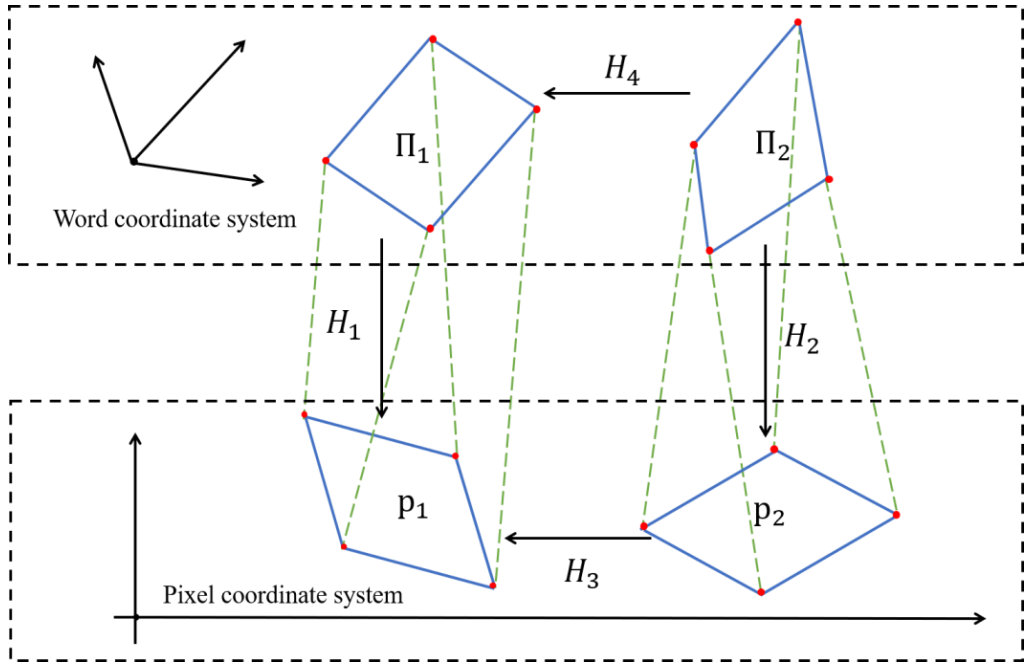

Figure 3: Homography transform among two planes in the space and their corresponding images at the pixel plane.

Notice that the transformation from $\Pi_2$ to $\Pi_1$ realizes a point-to-point correspondence, which means that there is no rotation and translation between the two planes. Therefore, the homography transform from $P_2$ to $P_1$ in the pixel plane is equivalent to the homography transform from $\Pi_2$ to $\Pi_1$ in space. Since the original datasets are all collected from the pixel plane, the homography transform at the pixel plane realizes the alignment of the data set and camera gesture correction of each scanning location.

## 2.3 Location error correction in the frequency domain

After the processing in the previous section, the original dataset can be aligned, nevertheless, the default step of the camera is not as same as the actual step, resulting in the non-corresponding of the spatial constraints and the frequency domain constraints. The reconstruction algorithm of camera-scanning Fourier ptychography relies on the spatial constraints from the dataset and the frequency domain constraints of the pupil function to find a convergent solution iteratively that satisfies both the constraints simultaneously. The purpose of location error correction is to find a particular location of the frequency domain constraint for each image from the data set. Since the camera aperture is located at the Fourier plane, the spatial location of the aperture linearly determines the location of pupil in the frequency domain during camera scanning. Our goal is to convert $(\Delta u, \Delta v)$ into the actual location offset in the frequency domain for accurate reconstruction. Figure 4 shows the transformation from pixel length of the image to pixel length in the frequency domain.

Defining the camera focal length $f$, camera aperture $D$, pixel size $\Delta$, wavelength $\lambda$, the optical resolution is

$$RES_{Aperture} = \frac{1.22 \lambda f}{D}. \tag{10}$$

The cut-off frequency is

$$K_{Aperture} = \frac{1}{2 \times RES_{Aperture}}. \tag{11}$$

The Nyquist frequency determined by the pixel size is

$$K_{Max} = \frac{1}{2 \times \Delta}. \tag{12}$$

The reconstruction process needs discretization. Let the pixel dimension is $M$ corresponding to $K_{Max}$. Then the size of the circular filter of the pupil function after discretization can be deduced as

$$D_{Pixel} = M \frac{K_{Aperture}}{K_{Max}}. \tag{13}$$

So, the ratio from the physical length of the Fourier plane to the pixel length in frequency domain is

$$Ratio_1 = \frac{D_{Pixel}}{D}. \tag{14}$$

Camera calibration establishes the transformation between the physical length of the sample plane and the pixel length of the image. Notice that the sample is fixed during the scanning process, therefore, the pixel drift is entirely caused by camera scanning. So, the ratio from the pixel length of the image to the physical length of the Fourier plane is equal to the ratio from the pixel length to the physical length of sample plane. Here, let the ratio be $Ratio_2$.

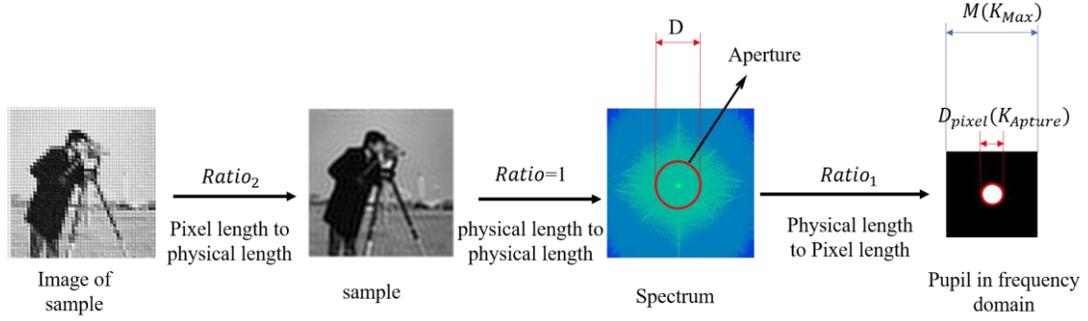

Figure 4: Transformation from pixel length of image to pixel length in frequency domain.

Using the location offset $(\Delta u_i, \Delta v_i)$, the actual location offset in the frequency domain of the $i^{th}$ scanning aperture is obtained as

$$\hat{\Delta u}_i = \Delta u_i \cdot Ratio_1 \cdot Ratio_2 \qquad (15)$$

$$\hat{\Delta v}_i = \Delta v_i \cdot Ratio_1 \cdot Ratio_2 . \qquad (16)$$

The corresponding intensity image of the $i^{th}$ scanning can be written as

$$I_i(u',v') = |\mathcal{F}^{-}\{O(u',v')P(u'-\hat{\Delta u}_i, v'-\hat{\Delta v}_i)\}|^2 \qquad (17)$$

where $O(u',v')$ represents the Fourier spectrum of the sample in the frequency domain. $P(u'-\hat{\Delta u}_i, v'-\hat{\Delta v}_i)$ is the pupil function with a center $(\hat{\Delta u}_i, \hat{\Delta v}_i)$. Take the central image of the brightfield as the reconstruction center, the reconstruction process of camera-scanning Fourier ptychography with location error correction is as following steps.

(1). The first step is to estimate the Fourier spectrum of the sample. This value can be arbitrary, and let $O(u',v') = 0$.

(2). The location offset $(\hat{\Delta u}_i, \hat{\Delta v}_i)$ is applied, the estimated value of the spectrum after the pupil is

$$\psi_i^k = O^k(u',v')P^k(u'-\hat{\Delta u}_i, v'-\hat{\Delta v}_i) \qquad (18)$$

Where $k$ represents the $k^{th}$ iteration, the wavefront distribution of the sensor is

$$\Phi_i^k = \mathcal{F}^{-}\{\psi_i^k\}. \qquad (19)$$

(3). Replacing the amplitude of $\Phi_i^k$ with the square root of the $i^{th}$ image $I_i$, we can get

$$\hat{\Phi}_i^k = \sqrt{I_i}\frac{\Phi_i^k}{|\Phi_i^k|}. \qquad (20)$$

(4). Implementing Fourier transform on $\hat{\Phi}_i^k$ to obtain $\hat{\psi}_i^k$, which is used to update the spectrum and pupil function

$$\hat{\psi}_i^k = \mathcal{F}\left\{\hat{\Phi}_i^k\right\}. \tag{21}$$

The update formulas of the spectrum and pupil function are as follows

$$O^{k+1}(u'-\hat{\Delta u}_i, v'-\hat{\Delta v}_i) = O^k(u'-\hat{\Delta u}_i, v'-\hat{\Delta v}_i) + \alpha \frac{P^{k*}(u',v')}{|P^k(u',v')|_{\max}^2}(\hat{\psi}_i^k - \psi_i^k) \tag{22}$$

$$P^{k+1}(u',v') = P^k(u'-\hat{\Delta u}_i, v'-\hat{\Delta v}_i) + \beta \frac{O^{k*}(u'-\hat{\Delta u}_i, v'-\hat{\Delta v}_i)}{|O^k(u'-\hat{\Delta u}_i, v'-\hat{\Delta v}_i)|_{\max}^2}(\hat{\psi}_i^k - \psi_i^k) \tag{23}$$

where $\alpha$ and $\beta$ are the search step of the algorithm.

(5). Repeat steps 2 to 4 until the dataset is traversed and the iteration converges.

## 3 Simulation

To verify the feasibility of the pose correction algorithm, we implement a simulation with a data set size of $15\times15$. When collecting the original data set, assuming the camera scans and records the LR images with a fixed step, and on this basis, random step error and twist error are added to each scanning. Except for the ideal situation, each scanning contains pose errors, which will make reconstruction results degraded. Using the principle of the extraction of location information described previously, the extraction results are shown in Figure 5. The simulation results show that the extraction errors of 90% of the scanning locations are less than 1.75 pixels, and the maximum extraction error is 2.97 pixels. The method can accurately extract the image scanned by the camera.

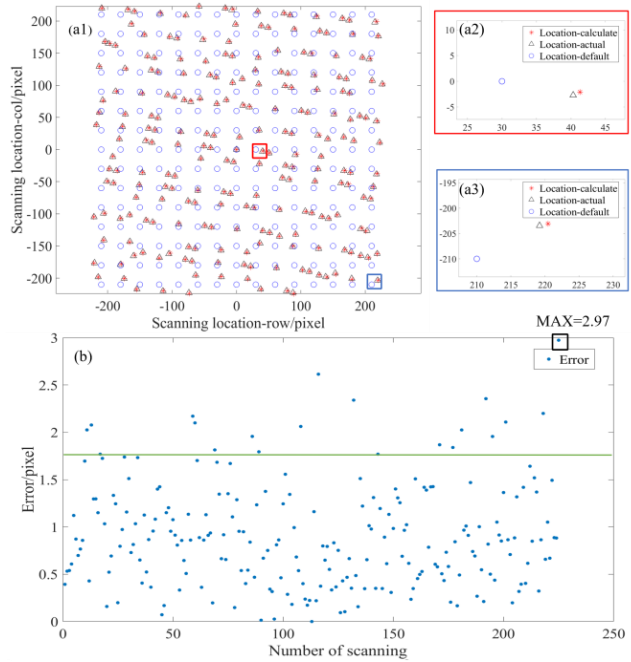

Figure 5: Experimental results of location information extraction using the mentioned method. Applying homography transform by using specific points, we can obtain a data set with camera pose error. The default pixel offset between adjacent scanning locations is 30 pixels, and the absolute value of the random step error is no more than 15 pixels. Besides, we add different pixel offsets with absolute values of no more than 2 pixels to specific pixel locations of a single image to simulate the twist of the camera. (a1) shows the actual camera location, the camera location extracted by the algorithm mentioned, and the default location when the camera scans at the specified step size. (a2), (a3) show the corresponding region of interest respectively. (b) Error of the extraction of each scanning location by calculating the distance between two points. The green line represents the standard of 1.75 pixels.

After the extraction of the location information. We implement a simulation of the reconstruction using the alternating projection algorithm. A comparison of the results of the simulation is shown in Figure 6, demonstrating that the pose correction method proposed in this paper can effectively improve the quality of reconstructed images.

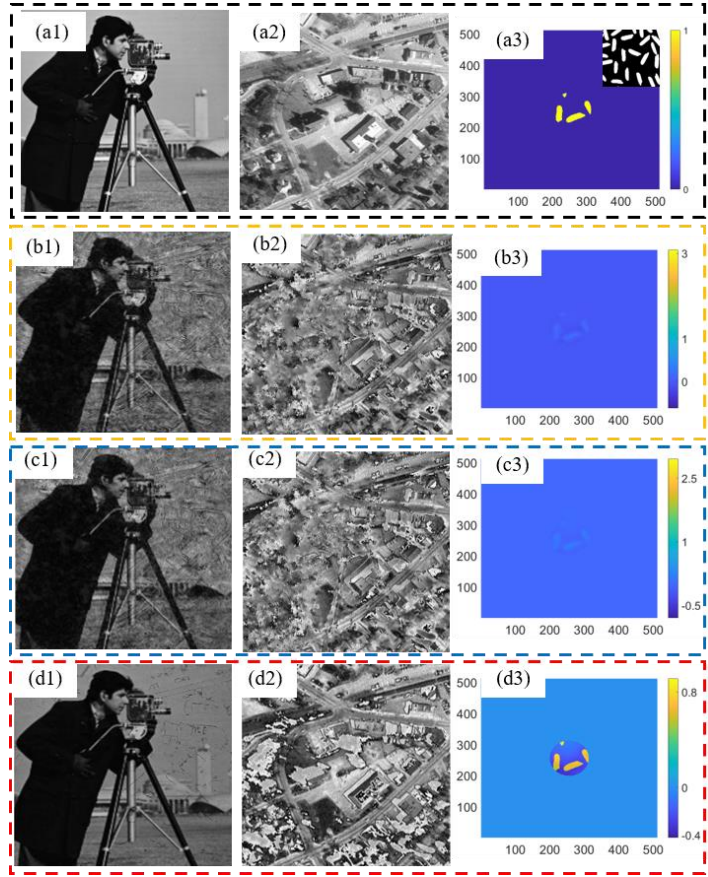

Figure 6: (a1), (a2), and (a3) respectively show the input high-resolution amplitude, phase, and pupil, we superimpose a rice image in the low-pass filter to simulate the pupil of the optical system. (1), (2), and (3) represent amplitude result, phase result, and pupil result respectively. (b) shows the reconstruction result of image alignment only using the location error (the reconstruction error caused by the change of camera gesture is not considered). (c) shows the reconstruction result of image alignment using the homography transform without location error correction in the frequency domain. (d) shows the reconstruction result using homography transform for image alignment and location error correction in the frequency domain.

## 4 Experiment

To verify the effectiveness of the proposed scheme, an indoor setup of camera-scanning Fourier ptychography was built, the system is shown as Figure 7.

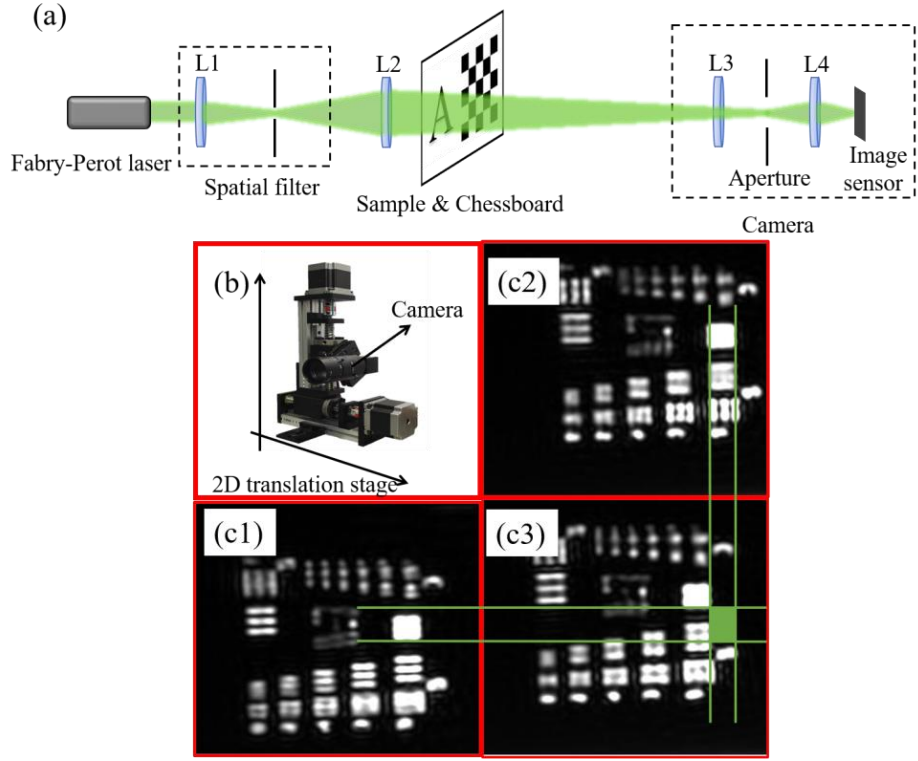

Figure 7: (a) The setup of the reported scheme. L1 is a lens with a focal length of 58mm, used to converge the beam. Since the Fraunhofer diffraction condition cannot be satisfied in the laboratory, a lens L2 is added with a focal length of 300mm to realize the Fourier transform of the sample [1], L3 and L4 are the imaging lenses. (b) The camera and the 2-D mechanical translation stage used in the experiment. (c1) shows the image located at [8,8] of the data set. (c2) shows the image located at [8,9]. (c1) shows the image located at [9,8]. (c1) and (c2) are adjacent to (c3) in different directions. We select specific positions in these images to compare the differences between the scanning process from (c3) to (c1) and (c3) to (c2). The green area is rectangular rather than square, demonstrating the appearance of the pose error as the camera scanning.

We use a Fabry-Perot laser with a wavelength of 520nm. A camera with a sensor pixel size of 2.2μm is located at the conjugate plane of the point light, coupled with a lens with a focal length of 75 mm ($F30$) mounted on a motorized 2D stage. We use a pinhole to filter out high-frequency noise to make the outgoing laser more uniform. The sample is placed at a distance of 83mm from the camera. A chessboard with corner points of 6×9 and a side length of 2mm is placed on the sample plane. During the camera scanning process, it is ensured that the sample and the chessboard are in the field of view at the same time.

In the first set of experiments, the USAF1951 resolution target was used as the sample to verify the effectiveness of the scheme. The acquired image array was 15×15 with a scanning step of 0.6 mm corresponding to an overlap rate of

74.4%. The experiment results are shown in Figure 8. Whether the amplitude result or the phase result, have been optimized obviously with the camera pose correction scheme proposed in this paper.

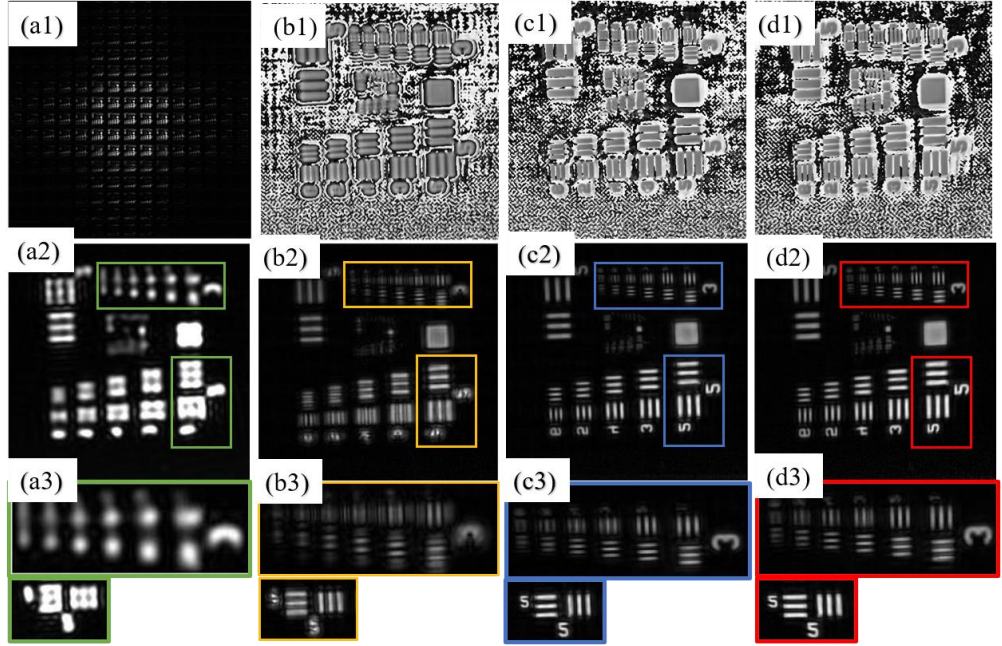

Figure 8: The comparison of the experimental results of USAF1951 target. (a1) The original data set. (a2) The central image of the brightfield. (a3) The corresponding region of interest in (a2). (b1), (b2), and (b3) respectively show the phase, amplitude, and the corresponding region of interest of the reconstruction result of image alignment only using the location error. (c1), (c2), and (c3) respectively show the phase, amplitude, and the corresponding region of interest of the reconstruction result with image alignment using the homography transform without location error correction in the frequency domain. (d1), (d2), and (d3) respectively show the phase, amplitude, and the corresponding region of interest of the reconstruction result using homography transform for image alignment and location error correction in the frequency domain.

In the second set of experiments, we use a bee wing sample. The acquired image array is $15 \times 15$ with a step of 0.7mm corresponding to an overlap rate of 70.1%. The experimental results are shown in Figure 9. Since the images at the junction of brightfield and darkfield will degrade the quality of reconstruction, such images are removed by setting a threshold during the reconstruction process. Generally, about 8 images are removed in a $15 \times 15$ array with an overlap of 70%. Compared with the reconstructed amplitude with pose correction, the details of the reconstructed amplitude without Pose correction contain several artifacts and blurs, resulting in distortion of reconstruction. By using the pose correction scheme described in this paper, the reconstructed results are greatly improved. Note that the quality of the phase image is also optimized. The black area in the phase image is caused by the thick edges of the bee wing specimen.

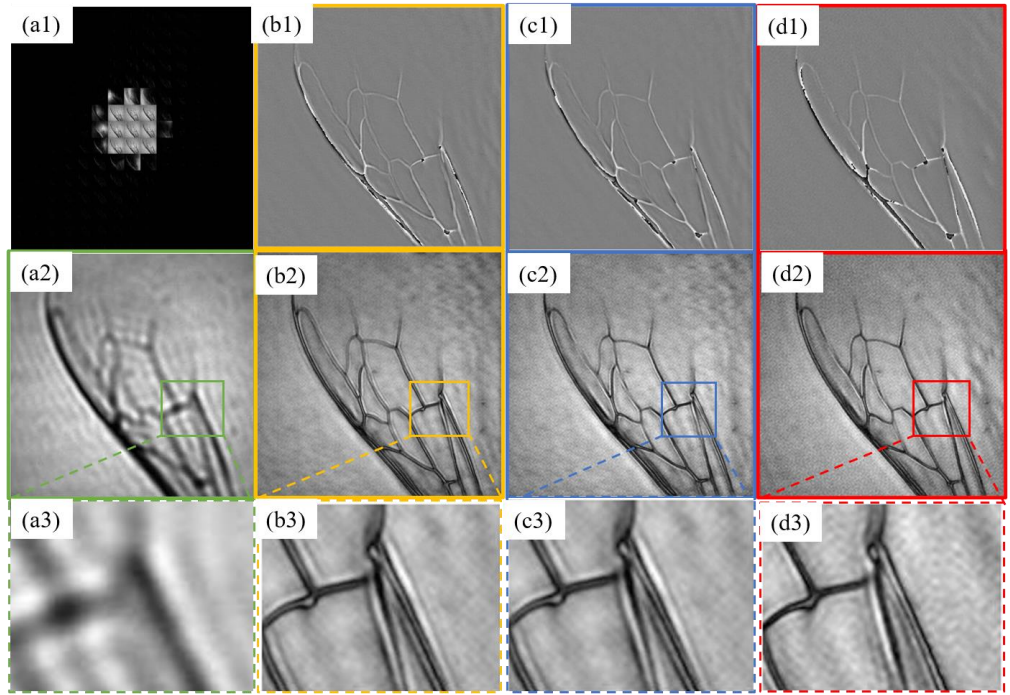

Figure 9: The comparison of the experimental results of the bee wing. (a1) The original data set. (a2) The central image of the brightfield. (a3) The corresponding region of interest in (a2). (b1), (b2), and (b3) respectively show the phase, amplitude, and the corresponding region of interest of the reconstruction result of image alignment only using the location error. (c1), (c2), and (c3) respectively show the phase, amplitude, and the corresponding region of interest of the reconstruction result with image alignment using the homography transform without location error correction in the frequency domain. (d1), (d2), and (d3) respectively show the phase, amplitude, and the corresponding region of interest of the reconstruction result using homography transform for image alignment and location error correction in the frequency domain.

## 5 Summary and discussion

Focusing on that the images of the data set have pixel offsets caused by the change of camera pose and the errors of camera scanning locations cause the incorrect constraints in the frequency domain during the reconstruction process in camera-scanning Fourier ptychography, we propose a camera pose correction scheme based on camera calibration and homography transform. We use the homography transform to align the dataset and use camera calibration to solve the relative pose between the sample and camera of each scanning location. we extract the location offset of the camera and impose it in the reconstruction process, and finally, the pose error of the camera is corrected. The simulation and experiment verified that the scheme can remove the artifacts caused by errors of constraint in spatial domain and frequency domain and realize high-quality camera-scanning Fourier ptychography. Combined with the algorithm for feature recognition, we can use features in the environment to implement camera calibration and homography transform to realize camera pose correction without adding features in the environment artificially. If using a camera array, we can use the baseline distance of cameras as constraints to calculate the camera parameters for camera calibration. Combined with reflective FP systems, which are more suitable for long-distance imaging, the scheme proposed in this paper provides the possibility for applying FP in remote sensing imaging and space imaging. These will be our next work consideration.